\documentclass[prb]{revtex4}
\usepackage{graphicx}

\begin{document}

\title{Static polarizability of molecular materials: \\
environmental and vibrational contributions}

\author{Francesca Terenziani, Anna Painelli}
\affiliation{Dipartimento di Chimica GIAF
 Universit\`{a} di Parma, 43100 Parma, INSTM UdR Parma, Italy}
\author{Zoltan G. Soos} 
\affiliation{Department of Chemistry, Princeton University,
Princeton, New Jersey 08544, USA}

\date{\today}

\vfill
\eject

\begin{abstract}

Modeling the dielectric behavior of molecular materials made up of large
$\pi$-conjugated molecules is an interesting and complex task. 
Here we address linear polarizabilities, and the related dielectric constant,
 of molecular crystals and aggregates made up
of closed-shell $\pi$-conjugated 
molecules with either a non-polar or largely polar ground-state, and 
also examine the behavior of mixed-valence (or charge-transfer) 
organic salts. We recognize important 
 collective phenomena due to supramolecular
interactions in materials with large molecular polarizabilities, and 
underline large vibrational contributions to the polarizability 
in materials with largely delocalized electrons.

\end{abstract}

\maketitle

 \section{Introduction}
 
A static electric field $F$ applied to an insulating
  material induces  a dipole,
 $\mu_{ind} = \alpha F + \beta F^2/2 +\gamma F^3/6+.. $,
 whose magnitude depends on the polarizability, $\alpha$, and
 hyperpolarizabilities, $\beta$, $\gamma$...,
 of the system. Large (hyper-)polarizabilities imply easy redistribution
 of electronic or nuclear charge by the field.
 For isolated molecules in the gas phase, the formal theory of static
 polarizabilities and hyperpolarizabilities is well known, and accurate
 results can be obtained from quantum chemical calculations on simple
 molecules.\cite{bishop_rev}
  For complex molecular structures the calculation becomes
 challenging, particularly in large $\pi-$conjugated molecules, where
 important contributions to (hyper-)polarizabilities are expected from
 molecular vibrations.\cite{painelli98}
  For molecules in solution the polarization of
 the solvent screens the applied fields and the calculation
 of (hyper-)polarizabilities becomes fairly involved.\cite{agren}
 
In the solid state, and particularly in crystals, the definition of
 the dipole moment itself is challenging, and a formal theory
 for the polarization, $P$, the dipole moment for unit cell,
  was developed only in the last decade \cite{resta}.
 The  problem is that
 the polarization of the material apparently depends on the
 choice of the unit cell.\cite{souza} This problem finds a natural
 solution in the so-called Berry-phase formulation of  $P$, that
 relates the macroscopic polarization of the crystal to the ground state
 (gs) wavefunction rather than to the gs charge-density.
 \cite{vanderbilt,ortiz,resta_rev,resta}
 A key concept in the modern theory of polarization is that $P$
 is by itself ill-defined and not accessible experimentally:
 only {\it variations} of $P$ can in fact be defined as independent
 of boundary-conditions, and are experimentally accessible.\cite{resta_rev}
 The Berry-phase formulation of polarizability in insulating
 materials constitutes an important advance in understanding
 the electrical properties of crystalline materials, but does not
 solve the problem of the definition of a dipole moment operator
 in systems with periodic boundary conditions.\cite{siv,rignanese}
 A complete and coherent
 picture  for the polarizability (and hyperpolarizability) of
 crystalline materials is still lacking.
 
The Berry-phase definition of $P$  is particularly
 important in covalent insulators, i.e. in crystals such as oxides 
or silicon, where the charge density is sizeable even between atoms. 
In organic molecular crystals, by contrast, the overlap between 
electronic wavefunctions on different molecules
is small and there is an unambiguous natural way of
 partitioning the crystal into an assembly of neutral, localized
 unit cells (the so-called Clausius-Mossotti limit).
 \cite{resta_rev,souza}
 In the zero overlap approximation for charge-densities on different molecules,
 the dipole moment operator for the crystal is simply the sum of
 the dipole moments of the basic unit cells, and standard approaches
 to linear and non-linear polarizabilities apply: the formal
 problem of calculating the crystal polarization and the relevant
 (hyper)polarizability is trivially solved in this limit.
 The actual calculation of the electrical responses of molecular
 crystals is difficult, however, particularly for crystals of highly
 polarizable molecules. In these materials in fact
 important collective effects are expected\cite{mcconnell} and, even
 if the dipole moment can be calculated as the sum of local
 dipole moments, the polarizability and the hyperpolarizabilities
 are not the sum of local contributions.
 
In this paper we address the calculation of the linear polarizability
 $\alpha$ in crystalline materials, as related to the experimentally accessible
 static dielectric constant. Sections 2 and 3 are
 devoted  to the analysis of organic molecular crystals with little
 or no intermolecular overlap.
 In these insulators the zero-overlap approximation for charge densities allows
 to adopt standard expressions for the polarizability tensor,
 whose Cartesian components $\alpha_{i,j}$ measure the curvature of the gs 
energy $E_0(F)$ with respect to the field:
 \begin{equation}
 \alpha_{ij}=-\left(\frac{\partial ^2 E_0(F)}{\partial F_i \partial F_j}
  \right)_0 = 2{\sum_R}^\prime \frac {\langle G |\mu_i|R\rangle \langle R|
 \mu_j|G\rangle}{E_R}
 \end{equation}
 The sum-over-state (SOS) expression
 is the second order correction. It involves the excited states $R$ at
 energy $E_R$ above the gs at $F=0$ and the dipole operator $\mu$. 
The first
 expression for $\alpha$ above is the finite-field (FF) result, that
 requires only the gs, albeit at finite fields.
 We emphasize the long-range nature of electrostatic
 interactions that lead to significant perturbations to $\alpha$
 as due to the environment even in the absence of direct
 intermolecular overlap.
 Section 2 reviews a recent approach to the detailed
 calculation of $\alpha$ in non-polar  chromophores
 with an extended and strongly polarizable $\pi$-system.
 This approach, based on the FF calculation of $\alpha$ for
 molecules experiencing the electric field generated by the surrounding
 molecules, relies on  quantum chemical models for the
 isolated molecular fragments and  leads to quantitative estimates of
 the dielectric constant for crystals and films that nicely compare
 with experimental data.
 
\begin{figure}
\begin{center}
\includegraphics* [scale=0.8] {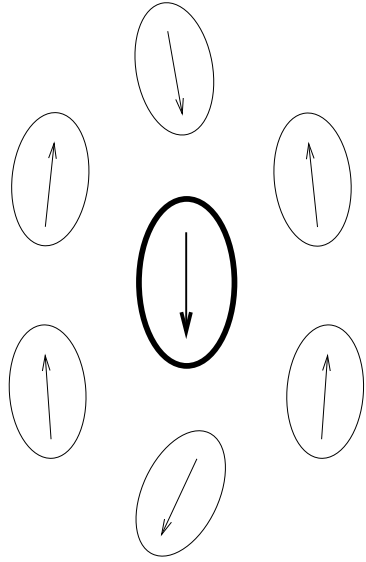}
\end{center}
\caption{Schematic view of 
dipolar molecule interacting with the surroundings, not drawn to 
minimize energy.  The arrows represent both dipoles and induced 
dipoles.}
\label{figura1}
\end{figure}

Section 3 presents an instructive toy model for clusters of polar and
 polarizable chromophores, based on a two-state picture for each chromophore.
 Whereas this simple model is hardly quantitative, it
 correctly grasps the basic physics of supramolecular interactions
 in these materials. Very schematically, Fig. \ref{figura1} depicts a polar 
chromophore in an environment of polar molecules that can be either an 
ordered crystal or a solution. Electrostatic interactions among dipoles 
and induced dipoles require a self-consistent treatment that accounts 
for large local fields due to the dipoles and induced dipoles themselves. 
Huge collective effects are possible in systems of polar-polarizable 
molecules, and are indeed expected for certain ranges of parameters. 
We note that the dipoles in Fig. 1 may represent both electronic and 
nuclear degrees of freedom or more complicated charge distributions. 
Quite generally, the zero-overlap approximation for molecular aggregates 
or crystals leads to such electrostatic problems.
 
Models with a limited number of states are then particularly
 instructive for assessing the inevitable approximations in molecular
 or extended systems.
 The Born-Oppenheimer (BO) separation of electronic and
 nuclear degrees of freedom is ubiquitous for molecules and generates
 potential energy surfaces (PES).
 In Section 4 we will assess the BO approximation for $\alpha$
  in models whose exact eigenstates
 are accessible, and will present a very simple expression for the vibrational
 contributions to $\alpha$ that applies in the limit of low vibrational
 frequencies.
  
 Based on these results, in Section
 5 we briefly address the role of vibrational
 degrees of freedom  in charge-transfer (CT) salts with a
 mixed (donor acceptor, DA) stack motif and variable ionicity.
 A first description of these materials was given in terms of
 non-overlapping DA pairs,\cite{soos78} so that collective effects in
 $\alpha$ are easily predicted. However, any realistic model
 for CT salts must account for charge delocalization along the stack. 
Since the Clausius-Mossotti approximation no longer applies, we resort 
to Berry-phase polarization and consider $P$ at instabilities. We relate 
the peak in the dielectric constant observed in some CT salts at 
a Peierls instability to large charge-fluxes induced along the stack 
by lattice vibrations.

\section{Electronic polarization in organic molecular crystals 
and thin films}

The polarization energy of a charge $q$ in a cavity of radius $a$
 in medium with dielectric constant $\kappa$ is $q^2(1  -1/\kappa)/2a$,
or $ \sim$ 1 eV
for typical $a \sim 5 $~\AA\  and $\kappa \sim3$ in organic molecular
crystals.\cite{gutmann}
The major role of polarization has long been appreciated in organic
semiconductors.\cite{pope,sili}
 The transport gap $E_t$ for creating an electron-hole
pair at infinite separation is reduced from the gas phase value by
$P = P_+ + P_-$, the separate stabilization of the cation and anion.
The major ($\sim $90\%) part of $P$ is thought to be electronic,\cite{sili}
with lattice relaxation or polarons accounting for the rest, and our
discussion of $P$ in this Section is restricted to electronic polarization.

We ask how the charge density $\rho(r)$ in the solid differs from the
gas-phase
  density $\rho_G(r)$ of molecules at the same nuclear positions.
The crystal induces the difference, $\Delta \rho(r) = \rho(r)  -\rho_G(r)$.
The lowest-order correction can again be viewed as polarization and
  rationalized by small overlaps in organic crystals. Organic molecular
  solids typically have van der Waals contacts and slightly shifted
electronic or vibrational excitations, as suggested by the
oriented gas  model that is the starting point of molecular exciton theory.
First-order corrections to energies clearly depend only on
$\rho_G(r)$, which can be inferred from gas-phase experiment or
theory. Electronic polarization is the first-order correction to
$\rho_G(r)$, or to the wavefunction, and is accordingly more difficult to
  compute.

  Small overlap at van der Waals separations suggests
starting with zero intermolecular overlap and introducing finite band
widths later. In the zero-overlap approximation, charge redistribution
$\Delta \rho(r)$ is clearly confined to molecules and intermolecular
interactions are purely electrostatic.\cite{tsiper}
 In this sense, the specific model
of polar-polarizable molecules introduced in the next Section represents any
aggregate. Conceptually, we have a crystal of quantum mechanical molecules
with classical interactions that may be viewed as a non-uniform electric
field $F(r)$. Since induced dipoles contribute to $F(r)$,
a self-consistent solution is required and made practical by
translational symmetry. For example, consider a crystal in a uniform
$F_{ext}$.
  Since all unit cells have the same total field, we apply constant
$F$ and resolve $F$ into local and external fields at the end.\cite{stp}
The calculation of $P_+$ or $P_-$ , the polarization energy of an isolated
charge, is done in two steps.\cite{tsiper}
 First, the {\it extensive}  quantity $\Delta \rho _N(r)$
  is found for the neutral lattice using translational
invariance. Then the {\it intensive} quantity $\Delta \rho_\pm(r,R)$
  is found relative to $\Delta \rho_N(r)$ in spheres of radius $R$
  about the charge without redistributing charge on molecules whose center
is outside the sphere. Convergence to $P_\pm$ rigorously goes as $1/R$
  at large $R$ with a slope that depends on the dielectric tensor $\kappa$.
Representative organic crystals show proper convergence by $R \sim 100$ \AA.
This corresponds to clusters of thousands of molecules and underscores
  the long-range nature of polarization. The polarization energy of CT
states also requires two steps, again starting with $\Delta \rho_N(r)$.
The sphere $R$ now encloses a cation and anion at fixed separation
  in the crystal lattice. The polarization energy converges faster, as
$1/R^3$, since a dipole is enclosed.

The zero-overlap approximation
reduces electronic polarization to finding $\alpha$ for molecules in a
non-uniform field $F(r)$. This major simplification yields a continuum
problem for the functional derivative,
  $\partial \rho(r)/\partial V(r')$, of the gs charge density at $r$
 with respect
to the  potential at $r'$. To be practical for crystals or
thin films of
  large molecules, we introduce a discrete approximation for
$\partial \rho/\partial V $ by restricting $r$ and $r'$
  at molecule $ a$ to the atomic positions $r_i^a$ and evaluating
$V(r)$ and $F =  -\nabla V$ at $r_i^a$.
  As done routinely in electrostatic (Madelung) calculations,\cite{metzger}
 we represent
the gas phase $\rho_G(r)$ by atomic charges $\rho_i^{ a(0)}$ at $r_i^a$.
Electronic polarization then yields new atomic charges and induced
atomic dipoles that are given by the following linear equations,\cite{tsiper}

\begin{eqnarray}
\rho_i^a &=& \rho_i^{a(0)} -\sum_j \Pi_{ij}^a\phi_j^a\\
\mu_i^a &=& \mu_i^{a(0)} -\tilde \alpha_i^a F_i^a
\end{eqnarray}
  Here the sum is over atoms $j$ of molecule $a$, the electrostatic
potential is $\phi_{i}^a = V(r_i^a)$, and $\Pi$ is the atom-atom
polarizability tensor,
\begin{equation}
\Pi_{ij} = - \left( \frac {\partial \rho_i}{\partial \phi_j }\right)_0
= - \left( \frac {\partial ^2 E}{\partial \phi_i \partial \phi_j}\right)_0
\end{equation}
that governs how charge redistributes within a molecule. The dipoles
$ \mu_i^a$ contribute to the potential and change according to
$F(r_i^a)$. The approximations of zero-overlap and discrete atomic
moments lead to eight linear equations per atom, namely the scalars
$\rho _i^a$, $\phi_i^a$ and the vectors $\mu_i^a$, $F_i^a$.
  They are solved iteratively starting with the oriented-gas potential
$\phi_i^{a(0)}$ produced by $\rho_i^{a(0)}$. Clusters of thousands
of molecules with about 50 atoms each are accessible on workstations.
\cite{tsiper,tsiper2}

We comment on general aspects of charge redistribution without
going into computational details that are found in refs. \onlinecite{tsiper}
 and  \onlinecite{tsiper2}.
First, semiempirical theory such as INDO/S \cite{indo}
 is particularly convenient for
$\Pi_{ij}$ since $\phi_i^a$ is then simply a site energy; this  
approximation, typical in solid-state models,
 is neither the 
 finite-field nor the SOS method of eq. (1),
but can be compared to them in exact models. Second, the best gas-phase
  polarizability $\alpha$ from experiment or theory is retained and
  partitioned between charge redistribution and a remainder,
the $\tilde \alpha_i^a$ above. An intuitive partitioning of $\alpha$
in acenes, as done in the submolecular method \cite{munn,sili}, leads to
self-consistent equations for induced dipoles in external
fields. Similarly, the gs dipole $\mu$ of polar
molecules can be partitioned between charge redistribution and the
$\mu_i^{a(0)}$ above. In the spirit of molecular exciton theory,
  we focus on how $\Delta \rho(r)$ changes in the actual crystal lattice
for molecules whose electronic structure is given by hypothesis.
Third, polarization energies are bilinear expressions \cite{tsiper}
 in self-consistent
  atomic charges and induced dipoles coupled to gas-phase potentials or,
vice versa, self-consistent potentials and fields coupled to
  gas-phase charges or dipoles. Since potentials and fields produced
by the best gas-phase charge distribution $\rho_G(r)$ can be evaluated
at atomic positions of the crystal lattice, first-order corrections to
polarization energies based on discrete $\rho_i^{a(0)}$
  can readily be found. \cite{sts} Such corrections are important in acenes or
other systems with electron-hole symmetry and hence
$\rho_i^{a(0)} \sim 0$ in the neutral molecule.
 
The optical dielectric tensor, or indices of refraction, of anthracene
crystals has been measured \cite{karl} separately and taken together yield the
principal values $\kappa_{ii}$ and uncertainties in Table 1.
The principal axes are the crystallographic $b$ axis,
by symmetry, and 
$\theta$ is the angle between $\kappa_{11}$ and $a$ in the $ac$ plane.
All calculations are based on the same, INDO/S-based $\Pi_{ij}$,
but have different molecular $\alpha$ inputs as indicated in the Table,
including the experimental polarizability. The triple-zeta basis
with field-induced polarization functions from ref. \onlinecite{reis}
and density functional theory (B3LYP) with a large basis are nearly
quantitative, while the oriented-gas value based on $\rho_G(r)$
  is clearly not. The dielectric tensor of the crystal includes charge
redistribution and induced dipoles whose interactions are treated
self-consistently via eqs (2) and (3).

\begin{figure}
\begin{center}
\includegraphics* [scale=0.8] {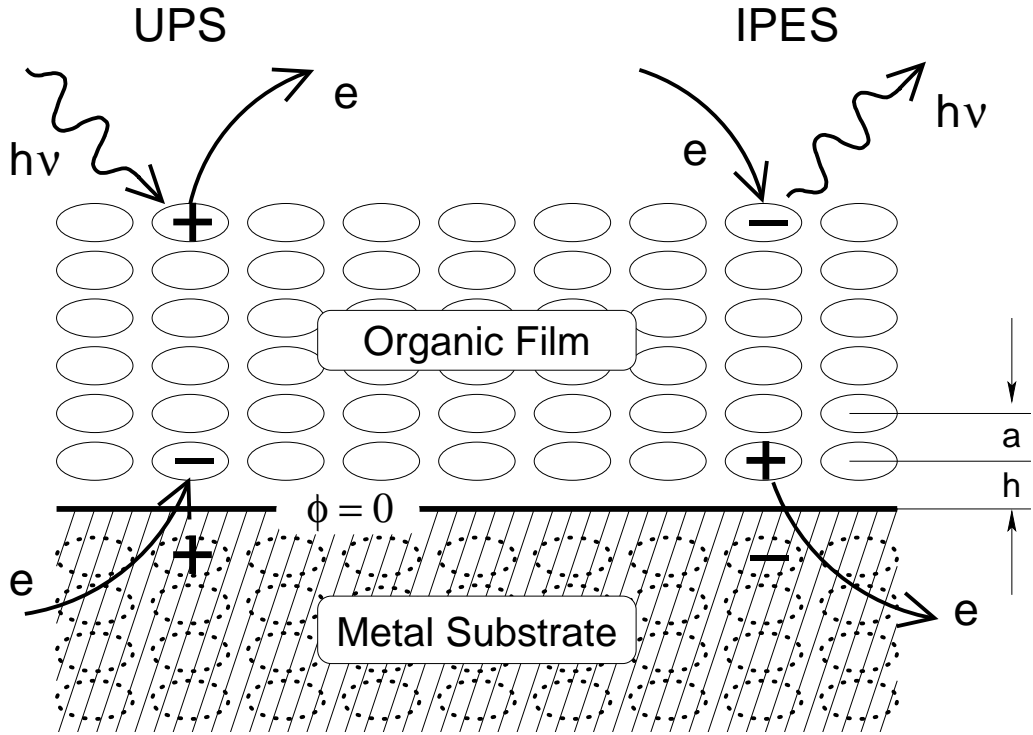}
\end{center}
\caption{Idealized model for electronic polarization in crystalline 
thin films on a metallic substrate at separation $h$. The $N$ layers
of the film appear as image charges. UPS and IPES generate a cation 
and anion at the surface, while charge injection generates ions
in the interface layer. }
\label{soosfig}
\end{figure}

Polarization energies $ P_+$ or $P_-$  are surface rather than bulk
measurements.\cite{salaneck}
 As sketched in Fig. \ref{soosfig}, photoelectron spectroscopy (UPS)
 involves $P_+$ for a cation that is mainly at the surface while inverse
photoelectron spectroscopy (IPES) involves $P_-$ . To minimize charging,
thin organic films on metallic substrates are used. Recent interest in
organic electronic devices is made possible by advances in forming and
characterizing crystalline thin films that compensate for the
limited mobility of charges.
  Charge injection then involves polarization at the metal-organic interface.
A constant potential surface with image charges is the simplest
model of the metal-organic interface. The zero-overlap procedure for
electronic polarization, eqs. (2) and (3), is readily extended to surfaces
or interfaces.\cite{tsgk} 
  We take $N$ molecular layers based on the crystal, place
them in van der Waals contact with the metal, and introduce image charges
and dipoles for neutral molecules as well as for ions at specified locations.
Instead of spheres of radius $R$ that enclose ions, we use pill-boxes of
thickness $2N$ and variable radius.
Convergence again requires thousands of molecules and can be monitored
with respect to $N$ as well as pill-box radius.

Perylenetetracarboxylic dianhydride (PTCDA) is a good hole conductor
and excellent film former, with molecules lying
almost flat on the surface.\cite{forrest}
 It is a prototypical molecule for organic devices.
The crystal value for $P = P_+ + P_-$  is 1.82 eV, which is in the
expected range.\cite{tsiper} The calculated monolayer value is
$P_{mono} $= 1.93 eV.\cite{tsgk}
  The large polarizability of image charges on one side offsets the vacuum
on the other side. Accordingly, thick films have reduced
$P_{surf}$ = 1.41 eV. The 500 meV change between monolayers and thick
films agrees quantitatively with UPS, IPES, and tunneling spectra of
PTCDA monolayers and films on gold and silver.\cite{tsgk}
 The calculated $P$
at the metal-organic interface, with image charges on one side and
crystal on the other, increases to $P_{inter}$ = 2.21 eV.
The separate values of $P_+$ or $P_-$  are relevant for matching energy
levels to facilitate the injection of holes or electrons. Efficient
injection is a major challenge whose pursuit is largely empirical at
present, usually without any consideration of polarization.

Pentacene is another widely used molecule, mainly as thin
film transistors.\cite{horowitz} Its herringbone structure is more common than
PTCDA stacks. Pentacene films have the long axis almost
normal to the surface, with high conductivity parallel
to the surface. The two inequivalent molecules per unit cell are
calculated \cite{tsiper2} to have slightly (70 meV) different $P$
  = 2.01 eV in the crystal. $P_{mono}$ differs by only 6 meV from the
crystal, while $P_{surf}$ is 0.23 eV less and $P_{inter}$
 is 0.13 eV greater.\cite{tsiper2}
  The contrasting structures of pentacene and PTCDA films lead to
different electronic polarizabilities that can now be estimated
in the well-defined limit of zero overlap rather than just expected
on general grounds in anisotropic solids. Similarly, accurate polarization
energies are needed for quantitative analyses of transport gaps or of CT
states seen in electroabsorption.

  In the context of polarizabilities, we
close this Section by noting that since the pentacene long axis is almost
normal to the metal, the images charges for a cation or anion produce
fields along the direction of largest $\alpha$ and redistribute charge to the
ends of the molecule. The induced dipoles of adjacent molecules
 are parallel and
close to each other. These repulsive interactions are relieved by
redistributing charge toward the molecule's center, as actually found
  in the self-consistent solution.\cite{tsiper2}
 The field $ F(r_i^a)$ at a pentacene
ion on a metallic surface vanishes for atoms near the center and
ranges from $\pm$10$^7$ V/cm for atoms close to and far from the metal.
Electronic polarization in organic molecular crystals produces
  highly non-uniform fields, as found 
explicitly in pentacene films and studied below in clusters of 
polar-polarizable molecules.

\section{Electronic polarization in clusters of polar-polarizable molecules}
\label{sec_agg}

Important intermolecular effects are expected in 
materials where several polar and highly polarizable molecules
interact. In these materials in fact the local fields generated by the polar
molecules strongly affect the charge distribution on the molecules
themselves, leading to a non-trivial self-consistent problem.
In so-called push-pull chromophores  an electron-donor (D) and an
electron-acceptor (A) group are linked by a $\pi$-conjugated
bridge. These intrinsically polar molecules are also highly
polarizable due to the presence of delocalized $\pi$-electrons.
They are also often largely hyper-polarizable, and represent
  the molecules of choice for second-order NLO applications.\cite{marder}
As originally recognized by Oudar and Chemla,\cite{oudar}
  the low-energy physics
of push-pull chromophores is well described by a two-state picture.
The gs resonates between
two limiting structures: $|DA\rangle$ and $|D^+A^-\rangle$, that can
be taken as the two basis states of a Mulliken model.\cite{mulliken}
These two states are separated by an energy $2z_0$ and are mixed by
a matrix element $-\sqrt{2}t$, that, with no loss of generality, will
be fixed as the energy unit. The relevant gs is written as
$|G\rangle=\sqrt{1-\rho}|DA\rangle+\sqrt{\rho}|D^+A^-\rangle$, where
the ionicity $\rho$ only depends on  $z_0$.
$\rho$ is proportional to the gs dipole moment:
$\mu_G=\mu_0\rho$ and $\mu_0=\langle D^+A^-|\hat\mu|D^+A^-\rangle$.
The excited state, $|E\rangle$, is orthogonal to $|G\rangle$ and its
dipole moment is $\mu_0(1-\rho)$. The excitation energy is
$\omega_{CT}=1/\sqrt{\rho(1-\rho)}$ and the corresponding
transition dipole moment is $\mu_0\sqrt{\rho(1-\rho)}$.

This simple model sets  the basis for current understanding of NLO
responses of push-pull chromophores and, if extended to account for
the coupling to molecular vibrations and solvation effects, it offers
a good description of the spectroscopic properties of push-pull
chromophores in solution.\cite{jpc1-2,baba}
  Here we adopt this model to investigate the
role of intermolecular interactions in clusters of polar-polarizable
chromophores.
We consider a cluster of Mulliken molecules
with purely electrostatic interactions.
The relevant Hamiltonian is:\cite{jacs}
\begin{equation}
{\mathcal H}_{int}= \sum_i(2z_0\hat\rho_i -\sqrt{2}t\hat\sigma_{x,i}) +
\sum_{i,j>i} V_{ij} \hat\rho_i\hat\rho_j
\label{ham_supra}
\end{equation}
The first term in (\ref{ham_supra}) describes the on-site problem, with
$\hat\rho_i$ measuring the polarity (i.e. the weight of the zwitterionic
state)
of the $i$-th chromophore, and $\sigma_{x/z,i}$ is the $x/z$-Pauli matrix
for the $i$-site. The second term accounts
for electrostatic intermolecular interactions with $V_{ij}$ measuring the
interaction between zwitterionic species located on sites $i$ and $j$.

\begin{figure}
\begin{center}
\includegraphics* [scale=0.8] {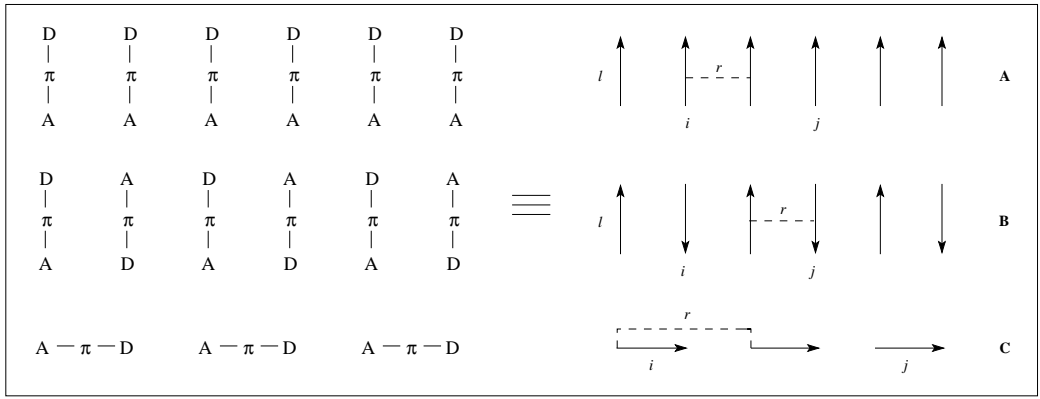}
\end{center}
\caption{Schematic view of the three one-dimensional clusters
of polar and polarizable molecules considered in this work, the left
panel shows the relative orientation of molecular dipole.}
\label{clusters}
\end{figure}

The Hamiltonian (\ref{ham_supra}) is fairly general. Here we consider
1-dimensional clusters of $N$ equivalent molecules with the three
geometries sketched in Fig.~\ref{clusters}.
We model each zwitterionic molecule as a segment of length $l$ carrying
$\pm e$ charges at the D/A ends, so that, for unscreened interactions,
$V_{ij}$ is fixed by $v=e^2/l$,
the interaction between two charges at unit distance,
and $r$, the  interchromophore distance.
In any case the specific expression
 for $V_{ij}$ does not alter the basic physics
of the model.
The above  Hamiltonian is easily written and diagonalized
on the $2^N$  basis obtained from the direct product of the two 
basis functions, $|DA\rangle$ and $|D^+A^-\rangle$, on each site.
By exploiting the translational symmetry we are able to find exactly
at least the lowest 30 eigenstates for systems with up to 16 sites.

\begin{figure}
\begin{center}
\includegraphics* [scale=0.5] {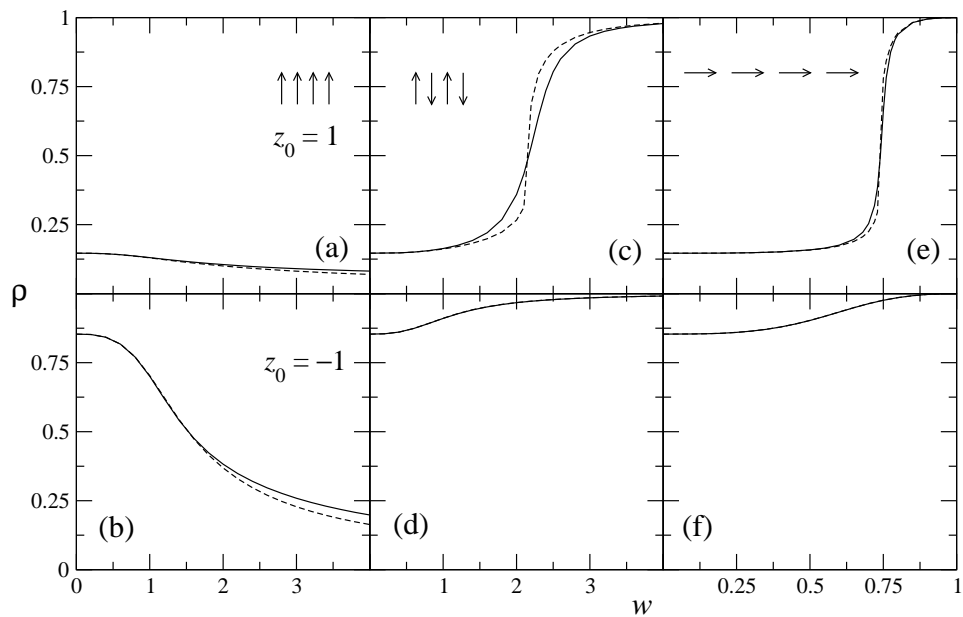}
\end{center}
\caption{Molecular ground state polarity, $\rho$, as a function of 
the strength of the interchromophore interaction, $w$, 
calculated for clusters of 16 molecules, with $v = 1$. 
Left, middle and right columns refer to geometries A, B and C, 
respectively; top and bottom rows correspond to $z_0 = 1$ and $-1$, 
respectively. Continuous and dashed lines refer to exact 
and mean-field results, respectively.
Finite size effects are negligible in all cases.}
\label{ro_mf}
\end{figure}

Fig.~\ref{ro_mf} shows the evolution of the  chromophore
polarity with the inverse interchromophore distance, $w=l/r$ for the three
lattices sketched in Fig.~\ref{clusters}.
All results are obtained for $v=1$; $z_0 $ is fixed to 1 in upper
panels, to show the behavior of a chromophore with a
  neutral (N) ground state in the gas phase
($\rho = 0.15$ at $w = 0$). The bottom panels ($z_0 =-1$) instead
describe the behavior of  zwitterionic (I) chromophores
($\rho =0.85$ at $w = 0$).
Interchromophore interactions disfavor charge separation in  A
geometry, and $\rho$ decreases with $w$ in the leftmost panels in
Fig.~\ref{ro_mf}, whereas just the opposite occurs for geometry B and
C (Fig.~\ref{ro_mf}, middle and right panels).
The behavior of an A cluster of I chromophores (Fig.~\ref{ro_mf}(b))
and of B and C clusters of N molecules (Fig.~\ref{ro_mf}(c) and (e),
respectively) are particularly interesting.
In the first case the isolated chromophore is zwitterionic, but,
with increasing $w$ (i.e. by decreasing the interchromophore
distance) the molecular polarity decreases down to the cyanine limit
($\rho = 0.5$) reaching the N regime for an interchromophore distance
of about 0.7 times the dipole length. Similarly, a neutral isolated
chromophore can be driven to the I regime for large enough
interactions in either B and C geometries when the interchromophore
distance is about one half of the dipole length (B cluster) or
about 1.4 times the dipole length (C cluster).

To make contact with experiment, we note that as a first estimate, 
push-pull chromophores have $\sqrt{2}t\sim 1$ eV. Then $v=e^2/l=1$
corresponds to  typical molecular lengths  ($l\sim 15$~\AA).
For these parameters, an  I
chromophore in A geometry crosses the I-N interface at $w\sim 1.5$
(panel (b)), i.e. for intermolecular distance of $\sim 10$~\AA,
a  reasonable situation. Similarly, curves in panels (c) and (e)
indicate that the N-I interface is crossed for interchromophore
distance of $\sim 7$~\AA\ and $\sim 20$~\AA\ for B and C cluster,
respectively. Again the inversion of the polarity occurs for
reasonable intermolecular distances.
  As a matter of fact, it has already been predicted
  \cite{reichardt} that the polarity
of a polar and polarizable molecule varies and eventually inverts due
to environmental interactions.
However, most of the approaches presented in the literature
on the properties of interacting push-pull chromophores disregard
the molecular polarizability and do not allow the molecular polarity
to readjust in response to supramolecular interactions.
On  general ground, push-pull chromophores have  large transition
dipole moments ($\sim$ 5-10~D) and  their permanent dipole moment
undergoes a large variation ($\sim$ 20-30~D) upon excitation \cite{baba}:
in samples with a medium-large
concentration of chromophores (intermolecular distances $\sim$ 5-10~\AA)
interchromophore interactions are a sizable fraction of, or even larger
than typical excitation energies ($\sim$ 1-3~eV) \cite{baba}.

The mf treatment of the previous Section becomes trivial when applied
to the above Hamiltonian.
In the two-state model in fact a single parameter, $\rho_i$, fully defines
both the molecular gs and local electric fields at the molecular positions.
For lattices of equivalent molecules, as in Fig.~\ref{clusters}, a
single-parameter self-consistent problem results from the mf approach.
In particular, within mf the lattice reduces  to a
 collection of non-interacting molecules, each one
described by the same two-state Hamiltonian as the isolated molecule,
but with a renormalized energy gap between $|DA\rangle$ and $|D^+A^-\rangle$,
  $z_0 \rightarrow z_0+m\rho$, where $m=\sum_{j}V_{ij}/2$.
Dashed lines in Fig.~\ref{ro_mf} show $\rho$ as obtained from the
self-consistent solution of the mf problem, and demonstrate that mf offers
a quite satisfactory description of the behavior of interacting molecules,
at least for not too large interactions.

Within mf it is easy to recognize a qualitative difference 
between the I to N crossover in repulsive lattices (A)
and the N to I crossover in attractive lattices (B and C). 
With increasing supramolecular interactions  in repulsive lattice
(A, $m>0$) the $\rho(z_0)$ curve becomes less negative, whereas it becomes more
negative in attractive lattices (B and C, $m<0$). For large negative $m$
($m<-2$)
 a divergent $\frac{\partial\rho}{\partial z_0}$ is expected, marking the
occurrence of a discontinuous crossover from the N to the I
regime.\cite{op2003} The N-I crossover is located at  $z_0 \sim -m/2$, and
for large $z_0$ ( $ > 1$) S-shaped $\rho(w)$ curves are calculated
within mf. The appearance of a discontinuous crossover
in mf treatments of C-lattices was discussed many years ago, and offered
a first description of the neutral-ionic phase transition observed in
CT crystals with a mixed stack motif.\cite{soos78} The behavior of the system
in the proximity of a discontinuous interface is very interesting, but
is beyond the scope of the present work.\cite{jacs}

\begin{figure}
\begin{center}
\includegraphics* [scale=0.7] {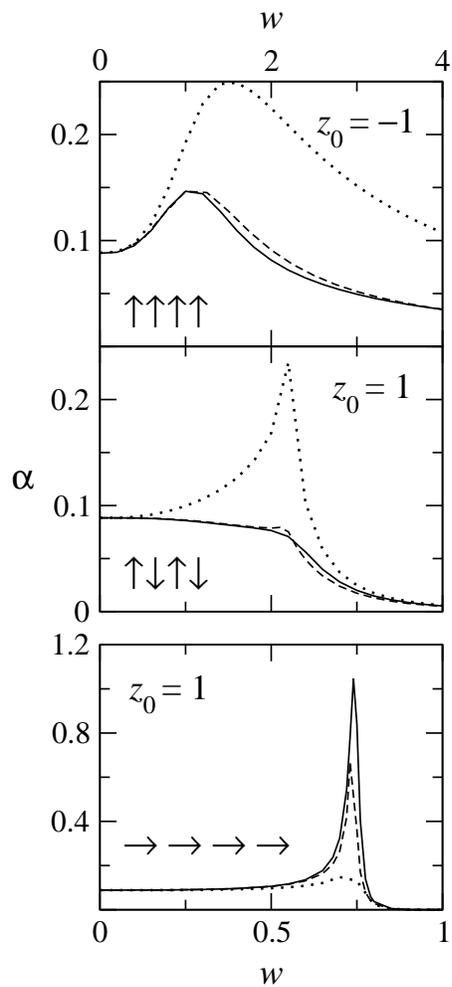}
\end{center}
\caption{Static polarizability 
calculated as a function of the strength of interchromophore
interaction, $w$, for clusters of 16 molecules, with $v=1$.
Top panel: A geometry, $z_0=-1$; middle panel: B geometry, $z_0=1$;
bottom panels: C geometry, $z_0=1$. Continuous lines refer to the exact
results, dashed lines to the mf-FF results, dotted lines to the 
mf-oriented gas results.}
\label{alpha_ff}
\end{figure}

The static susceptibilities  of systems described by the Hamiltonian in
Eq.~(\ref{ham_supra}) are easily obtained from the
successive derivatives
of  the ground-state dipole moment on a static applied field.
Since molecular dipole moments, in the proposed toy model, rigorously
lie along the molecular axis (say $z$), a single component of the
polarizability tensor ($\alpha_{zz}$) is relevant. The magnitude of this
component is shown in  Fig.~\ref{alpha_ff} for parameters corresponding
to panels b, c and e in Fig.~\ref{ro_mf}.
Continuous lines show exact results and demonstrate that
supramolecular interactions non-trivially affect
the molecular response.

To clarify the subtle physics governing
the responses of molecular clusters, we
discuss approximate approaches to the problem.
In the simplest and most widely adopted approach 
the response of a collection of chromophores is
calculated  as the sum of the responses of a collection of non-interacting
molecules with the same geometrical arrangement.\cite{dalton1}
 This {\it oriented gas} approach is 
however limited to very weak interactions and fails otherwise, since 
it completely disregards the dependence of $\rho$ on supramolecular 
interactions.
A slightly better approach relies  again on  the oriented gas
approximation, but assigns  each chromophore the same $\rho$ as 
obtained within the mf approximation.
The mf-oriented gas estimates of static susceptibilities are reported in
Fig.~\ref{alpha_ff} as dotted lines. The strikingly large deviations 
from exact results are quite unexpected: the gs polarity 
is fairly accurately calculated within mf, for these parameters. The 
failure of the mf-oriented-gas approximation appears since 
the response to an applied field of a molecule in the cluster 
differs from the response of an isolated molecule to the same 
field.
Much  better results can be obtained within  mf
provided the gs dipole moment of each molecule in the cluster
is allowed to readjust to the applied
field. A proper FF-mf calculation of the polarizability as the first 
derivative of the cluster dipole moment on the applied field just represents 
the (trivial) implementation to the Hamiltonian ~\ref{ham_supra} of
the self-consistent treatment described the Section 2.
As shown by dashed lines in Fig.~\ref{alpha_ff}, this approach nicely
compares with exact results, apart from deviations observed in a narrow
region around the N-I crossover.

The static linear polarizability is a gs
property that can be accurately calculated within mf,
provided collective behavior is properly accounted for.
Non-linear responses can be obtained within the same approach
from  the  successive
derivatives of the gs dipole moment, and collective effects 
are found to rapidly
increase with the order of non-linearity.\cite{op2003}
The deviations between dotted and dashed lines in Fig.~\ref{alpha_ff}
  demonstrate the
importance of collective behavior, here due to
the non-linear response of polarizable chromophores
to the perturbation induced by the surrounding.
Material properties are significantly affected by the
supramolecular arrangement:
linear and non-linear optical properties can be
strongly depressed or amplified by tuning intermolecular distances and/or
by changing the relative orientation of chromophores, with effects that
are more pronounced at intermediate polarities.
The design of molecular materials for advanced applications is then a
challenging task: the material properties in fact must be optimized at
the supramolecular level. The presented model just represents a first step
towards {\it supramolecular}
structure-properties relationships.

\section{Vibrational contributions to $\alpha$: a toy-model approach}

Both molecular vibrations and lattice phonons
are strongly coupled to delocalized electrons and can
significantly contribute  to susceptibilities.
Accounting for vibrational degrees of freedom even within a simple model
for molecular materials like that presented in the previous
Section is non-trivial. Exact non-adiabatic solution of the relevant problem
is already computationally demanding for clusters of just two
molecules.
In the following we therefore shortly discuss
the vibrational contribution to the polarizability of an isolated push-pull
molecule.
We again describe the chromophore as a Mulliken DA pair
(cf Section 3) but also account for Holstein
coupling  to a vibrational coordinate, $Q$.\cite{painelli98}
The relevant Hamiltonian is ($\hbar =1$ and $\sqrt{2} t=1$):
\begin{equation}
{\mathcal{H}} = 2z_0\hat{\rho}-\hat{\sigma}_x+
\frac{1}{2}\left(\omega^2Q^2+P^2\right)
-\sqrt{2\epsilon_{sp}}\omega Q\hat{\rho}
\label{hamiltonian}
\end{equation}
The first two terms in the above equation describe the electronic Hamiltonian
(cf previous Section).
In the third and fourth terms the vibrational problem is defined in terms
of   an internal vibrational coordinate, $Q$, and of its conjugated 
momentum, $P$.
In particular the two basis states are assigned two harmonic PES
with equal frequency ($\omega$) but displaced minima to account
for linear e-ph coupling, whose strength is measured by
  $\epsilon_{sp}$, the relaxation energy of $|D^+A^-\rangle$.
Whereas the model can quite easily be extended to account for quadratic
coupling (i.e. for different vibrational frequencies in the two basis states),
this adds an additional parameter to the model, without affecting
the basic physics. \cite{bibbia}.

The exact non-adiabatic eigenstates of the above Hamiltonian
are obtained by the 
 numerical diagonalization of the relevant 
matrix written on the basis of the
direct product of the two electronic
states $|DA\rangle$ and $|D^+A^-\rangle$, and of the reference
vibrational states (i.e. the eigenstates of the
harmonic oscillator in the third term of
Eq.~(\ref{hamiltonian})).\cite{delfreo}
The  basis is truncated by fixing a maximum number
of phonon states, $M$; the corresponding $2M\times 2M$ matrix  can be
diagonalized  up to  fairly large $M$ values, yielding numerically exact
non-adiabatic eigenstates. The minimum $M$  required
to get convergence depends on the model parameters and on
the  properties of interest. 

\begin{figure}
\begin{center}
\includegraphics* [scale=1.0] {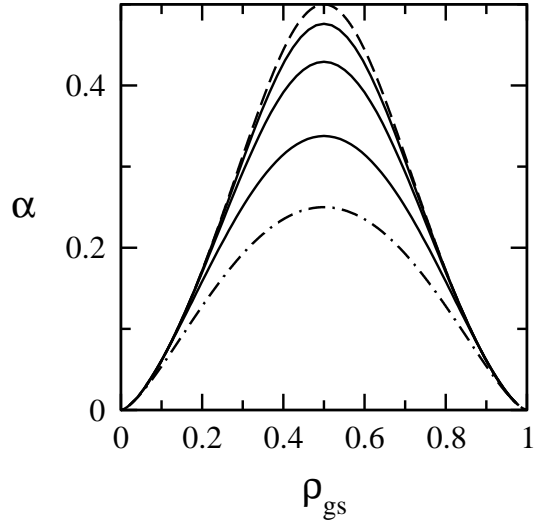}
\end{center}
 \caption{Static polarizability as a function of the ground state
ionicity $\rho_{gs}$, 
for $\epsilon_{sp} =1$ and different $\omega$ values ($\sqrt{2}t$ units).
Dashed line: potential energy contribution,
corresponding to $\omega = 0$ limit (see text);
dot-dashed line: bare electronic polarizability, corresponding
to the $\omega\rightarrow \infty$ limit (see text);
continuous lines report the exact polarizability calculated for 
$\omega=0.05$, 0.2, 1.0, 
smoothly evolving from the $\omega= 0$ to the
$\omega \rightarrow  \infty$ limits;
dotted line reports the polarizability calculated 
in the best harmonic approximation (see text).
Dotted and dashed lines are exactly superimposed. 
For the calculation 
dipole moments have been expressed in $\mu_0$ units.}
\label{alpha_omega}
 \end{figure}

The  second derivative of the exact
non-adiabatic gs energy vs the applied field, gives the exact
estimate of the molecular the polarizability. Results obtained for a
molecule with  $\epsilon_{sp} =1$ and a few $\omega$ values are
reported in Fig.~\ref{alpha_omega} (continuous lines). $z_0$ is tuned in this
calculation as to span the whole $0<\rho <1$ interval.
The dot-dashed line corresponds to the bare electronic
susceptibility, $\alpha_0$, i.e. to the response of the two-state model
with no e-ph coupling. Hence differences between the continuous lines and
the dot-dashed line  measure the vibrational contribution
to the static response.\cite{jcp}

For actual  molecules or complexes, the large number of electronic states
makes non-adiabatic calculations very demanding,
and the Born-Oppenheimer (BO) approximation is usually invoked.
In the BO approximation, the effective electronic Hamiltonian,
${\mathcal{H}}_{el}={\mathcal{H}}-P^2/2$, is defined by
subtracting the nuclear kinetic energy (KE)
from the total Hamiltonian. 
Its diagonalization yields  
analytical expressions for the ground and excited state potential 
energy surfaces (PES).\cite{jcp} 
Even if the Hamiltonian in Eq. \ref{hamiltonian} assigns the two basis 
states two harmonic PES with equal curvature, the PES for the ground
and excited state obtained from the diagonalization of the
electronic Hamiltonian have different curvatures and are largely
anharmonic.\cite{delfreo}
The anharmonicity
of the potential prevents the analytical solution of the vibrational
problem on either the ground or excited state; however, numerically exact
vibrational states can be calculated in both manifolds.\cite{jcp}
In particular,
the eigenstates of the harmonic oscillator with frequency $\omega$,
centered at the relevant equilibrium position, are a good basis
for the vibrational problem on either PES. The corresponding
  vibrational Hamiltonian
is the sum of a KE term, whose matrix elements are trivial in the
adopted basis, plus a PE term, whose matrix elements are calculated
via numerical integration.
Of course the vibrational matrix is diagonalized on a  basis
truncated to a   large enough number of phonon states as to
get convergence.

\begin{figure}
\begin{center}
\includegraphics* [scale=1.0] {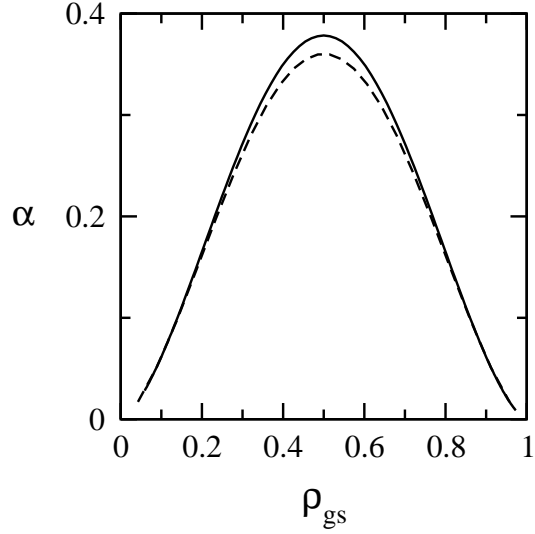}
\end{center}
\caption{Static polarizability as a function of the ground state
ionicity $\rho_{gs}$, for $\epsilon_{sp} =1$ and $\omega=0.5$ 
($\sqrt{2}t$ units).
The continuous line shows the non-adiabatic result; 
the dashed line shows the BO result.}
\label{alpha_nad}
\end{figure}

Once BO eigenstates are
obtained, the (transition) dipole moments entering the SOS expression
can be calculated via  numerical integration.
The static polarizability  calculated within BO approximation is
indistinguishable (in the scale of Fig.~\ref{alpha_omega})
from the exact one, as long as $\omega\leq 0.2$.
Fig.~\ref{alpha_nad} compares non-adiabatic (continuous line)
and BO (dashed line)
static susceptibility  for $\omega =0.5$, where deviations appear.
Of course the BO approximation becomes worse with increasing $\omega$ and
is totally untenable for $\omega\geq 1$. For push-pull
chromophores
$\epsilon_{sp}\sim\sqrt{2}t$  and typical vibrational frequencies,
$\omega\sim 1000$~cm$^{-1}<< \sqrt{2}t \sim 1$ eV  support the
validity of BO.

When applied within a FF approach to susceptibilities, the BO
approximation immediately leads to a partitioning of the
susceptibility into PE and KE contributions.\cite{jcp}
Within BO, the lowest eigenstate of the electronic Hamiltonian
defines the
PE for the motion of nuclei (in the gs manifold, of course).
The total gs energy is obtained by adding the nuclear KE to the PE.
Then, as long as BO applies, susceptibilities, i.e. the successive derivatives
of the gs energy with respect to an applied electric field, can be
calculated as sums of PE and KE $F$-derivatives.
The nuclear KE vanishes in the $\omega = 0$ limit, and the
($\omega$-independent)  PE polarizability, reported as the dashed line
in Fig.~\ref{alpha_omega}, represents the zero-frequency limit
of the exact polarizability.
The electronic polarization of the Holstein DA molecule
 is analytical  \cite{painelli98} and is reported as
the dot-dashed line in Fig.~\ref{alpha_omega}.
We can now understand the evolution of the vibrational contribution to the
polarizability with the vibrational frequency:
in the low-$\omega$ limit, KE contributions vanish, and the exact curve
tends to the limiting PE result. Vibrational contributions
are very large in this limit.
With increasing $\omega$, the vibrational contributions to the
response decrease: in the antiadiabatic limit ($\omega \rightarrow
\infty$) the vibrational contributions to the static polarizability
vanishes, and the  exact curve trivially reproduces the
bare electronic response.

The calculation of PE-susceptibilities for the toy-model in
Eq.~(\ref{hamiltonian}) is trivial. More generally it is easily implemented 
in quantum chemistry calculations and in solid-state models as well, since
it  only requires the gs energy calculated at the
relaxed geometry for different values of an externally applied field.
The calculation of KE contribution is more difficult,
since the $F$-dependence of the lowest vibrational state in the anharmonic
gs PES is needed.
The nuclear KE contributes to susceptibilities in two different
ways.\cite{jcp} First of all, due to anharmonicity, the molecular geometry
in the vibronic gs is different from the equilibrium geometry (corresponding
to the minimum of the gs PES).  This correction is
however very small.
The second contribution stems from the $F$-dependence of the nuclear
KE itself: it is this contribution that indeed accounts
for the deviations of the exact curves from the (dashed) PE curve (at least
in the BO regime, $\omega \le 0.2$, where non-adiabatic corrections are
negligible).
KE contributions  are of course very small for low $\omega$,
but they  increase with  increasing
$\omega$, leading to a suppression  of the
vibrational amplification of the static polarizability.
This is by no means accidental: in the antiadiabatic limit
($\omega\gg \sqrt{2}t$) phonons cannot contribute to the
static polarizability and, with increasing $\omega$,
KE contributions progressively
increase to counterbalance the PE contribution.

KE contributions to the linear polarizability exactly vanish
in the harmonic approximation.\cite{jcp}
In fact, the equilibrium position in any harmonic vibrational state
coincides with the bottom of the PES; moreover the nuclear KE is proportional
to the harmonic frequency, i.e. to the curvature of the PES. For a parabolic
PES, this quantity is obviously independent of $Q$, and hence of $F$.
Then, in the low $\omega$ limit, where KE contributions to susceptibilities
are negligible, the exact $\alpha$ (continuous line in Fig. \ref{alpha_omega})
coincides with the corresponding best-harmonic estimate, i.e. with the
estimate
obtained by modeling the gs PES as the parabola with the exact curvature
at the equilibrium. This is not true for hyper-polarizabilities
that, in the same $\omega \rightarrow 0$ limit, are strongly amplified
by the anharmonicity of the gs PES.\cite{jcp}
Within the best  harmonic approximation,\cite{delfreo}
$\alpha$ is conveniently partitioned into an electronic and vibrational
contribution:
\begin{equation}
\alpha=2\left[ \frac{\mu_{CT}^2}{\omega_{CT}}+\frac{\mu_{IR}^2}{\Omega}\right]
\label{aelavib}
\end{equation}
where $\mu_{IR}$ is the infrared transition dipole moment and
$\Omega$ is the frequency of the best harmonic frequency, 
i.e. the curvature of the gs PES at equilibrium.
The vibrational contribution to the linear polarizability is then
proportional to the infrared intensity of the coupled mode, at least
in the low-frequency regime, where PE contributions dominate the response.
We underline that this result only relies on the BO approximation
and on the neglect of KE contributions to the polarizability.
It therefore applies in the low-$\omega$ limit quite irrespective of 
the detailed model for the electron-phonon coupling, and also for
largely anharmonic PES.

 \section{Dielectric constant of CT salts}
 
Charge-transfer (CT) crystals have mixed face-to-face stacks of planar
 $\pi$-electron donors and acceptors as sketched in Fig. \ref{stack}.
 Intermolecular overlap is negligible between stacks, but not within 
stacks where $\pi-\pi$ overlap is indicated by less than van der Waals 
separation between D and A. The gs consequently has fractional 
charges $\rho$ at D and -$\rho$ at A sites. \cite{soosklein} 
Just as an example, in the prototypical material, TTF-CA, 
at ambient conditions about 0.2 electrons are transferred on average 
from the donor (tetrathiafulvalene, TTF)  to the acceptor 
(chloranil, CA).\cite{girlandonit} As seen in Fig. \ref{stack}, a 
regular stack of centrosymmetric molecules is not polar because there 
is an inversion center at each site, as in fact occurs in the actual 
structures. The inversion center is lost on dimerization and the 
gs becomes ferroelectric if dimerization is in the same sense everywhere.
 Much as it occurs in attractive lattices in Section 3, Madelung interactions
 favor charge separation, and
 a large variation of $\rho$ can be induced by tuning intermolecular distances.
 At $\sim$ 77 K TTF-CA undergoes a discontinuous
 phase transition to an I phase with $\rho \sim$ 0.7.\cite{girlandonit}
 Other systems with N-I transitions are known and transitions can be
 induced by temperature, pressure or by absorption of light. They are
 a complex and interesting phenomenon:\cite{tokurascience}
  both continuous and discontinuous
 transitions are known and, in all cases,  stack dimerization accompanies
 the charge crossover proving the important role of phonons
 and of e-ph coupling in these systems.

A sharp peak in the dielectric constant has been observed at the N-I
 transition of several CT crystals,\cite{tokurajacs}.
 It can be understood on general grounds as due
 to large charge fluxes induced by an applied
 field near the charge crossover, but microscopic modeling of dielectric 
peaks is still in progress.\cite{inprogress}
 Charges are delocalized along the stack and the zero-overlap 
approximation does not apply: we need a model for polarization 
in extended and highly correlated systems. Moreover the lattice 
(Peierls) phonon that induces dimerization
 is strongly coupled to electronic degrees of freedom in a manner 
reminiscent of solitons in polyacetylene. \cite{heeger}
 The Peierls mode in fact induces large charge displacements around 
the N-I crossover, and vibrational
 contributions to the polarizability cannot be disregarded.
 
The electronic structure of mixed stack CT salts can be described in terms
 of a  Hubbard model with only on-site electron-electron repulsion
 ($U$)  explicitly accounted for, and modified to account for the
 alternation of on-site energies, $\Delta$.\cite{anu}
  This model accounts only implicitly for
 Madelung interactions and hence describes only continuous
 N-I transitions.
 We also account for the
 coupling to the Peierls phonon, described by $\delta$, as follows:
 \cite{prl}
 
\begin{equation}
 H=-\sum_{i,\sigma} [1+(-1)^i\delta](c_{i,\sigma}^\dagger c_{i+1,\sigma}
+H.c.) +
 U \sum_{i}  \hat n_{i,\sigma} \hat n_{i,\sigma '}+
 \Delta \sum_{i,\sigma} (-1)^i c_{i,\sigma}^\dagger c_{i,\sigma}+
 \frac{N}{2\epsilon_d}\delta ^2
 \label{hamct}
 \end{equation}
 where $c_{i,\sigma}^\dagger$ creates an electron with spin $\sigma$
 on the $i$-th site and $\hat n_{i,\sigma} =c_{i,\sigma}^\dagger c_{i,\sigma}$.
 The last term in the above Hamiltonian measures the bare elastic
 energy associated with the dimerization mode, with
 $1/\epsilon_d$ measuring the lattice stiffness. 
The rigid lattice has $\epsilon_d$ = 0.
 
For $\Delta>>U$ the Hamiltonian (9) describes an almost
 N lattice ($\rho \rightarrow 0$) of donors and acceptors, 
whereas for  $U>>\Delta$ an almost I lattice of  spin 1/2 radical 
ions is obtained ($\rho \rightarrow 1$).\cite{anu} The two phases are
 qualitatively different, with the I lattice being unconditionally
 unstable to dimerization (spin-Peierls transitions, \cite{bray}). 
The N-I crossover can consequently be identified precisely even for 
continuous $\rho$ in the rigid regular stack.\cite{anu} 
Conditional instability in the N
 regime implies that soft lattices dimerize on that side
 before reaching the N-I crossover.\cite{prl} The Peierls
 phonon induces large charge fluxes along the chain with effects that
 increase the nearer the dimerization transition is to the N-I crossover of
 the rigid lattice.
 
By exploiting the recent definition of polarization in extended 
systems \cite{resta}, we were
 able to demonstrate huge peaks in the IR intensity of the Peierls mode
 at the structural instability.\cite{prl} But, as shown in the
 previous Section for Holstein modes, the IR intensity of vibrational 
modes is quite naturally
related to the system's polarizability. We draw two inferences: first, peaks
 in the static dielectric constant near the neutral-ionic phase
transition have a large contribution from
vibrational degrees of freedom; second such peaks in soft lattices 
are associated with the Peierls transition and occur at $\rho <$ 1/2. 
We anticipate that the Hamiltonian (9) 
or a closely related model will describe dielectric peaks.

 Since lattice phonons have low frequencies
 ($<$ 100 cm$^{-1} << t\sim$ 0.2 eV), the BO approximation applies and
 the KE contribution to $\alpha$  can be safely neglected.
 The calculation of the vibrational contribution to $\alpha$ in
 Eq.~(\ref{aelavib}) is simple, given the IR intensity of
 the relevant model. But the electronic contribution to $\alpha$
 is difficult for correlated electrons and will 
be presented separately.\cite{inprogress}
 
As a first estimate of the relative importance of the electronic 
and vibrational contributions, we consider the $U=0$ limit 
of uncorrelated electrons in (9), whose 
electronic $\alpha$ has long been known. \cite{cojan}   
 $U=0$ results are reported in Fig. \ref{epsilon} for 
$\epsilon_d=0.28$, which is a fairly stiff lattice that 
approximates TTF-CA. The static dielectric constant 
is related to polarizability by the standard expression:  $\kappa= 1
+(\alpha_{el}+\alpha_{vib})/\kappa_0$, where $\kappa_0$ 
is the permittivity of vacuum and $\alpha$ is the polarizability 
per unit cell. We took TTF-CA lattice parameters as typical for CT crystals. 
Moreover, for the sake of simplicity, we neglected 
the softening of the Peierls mode at the transition, by simply substituting 
the $\Omega$ entering the expression for the vibrational
polarizability in Eq.~(\ref{aelavib}) with the reference frequency
 ($1/\epsilon_d$). Thus the estimated vibrational
contribution to $\kappa$ is a lower limit.
 
The magnitude of electronic and vibrational contributions to $\kappa$ 
is similar for noninteracting electrons. Yet the peak at the
Peierls instability 
in Fig. \ref{epsilon} clearly has a vibrational origin. The physical 
basis of this
 behavior lies in the large charge fluctuations generated by Peierls phonons
 in the vicinity of a charge instability, and underlines once more
 the importance of properly accounting for e-ph coupling in systems
with delocalized electrons. 
All material properties, 
including {\it electronic} responses are in fact 
non-trivially affected by coupling to nuclear motions.
The electronic contribution, by contrast, 
is dominated in $U=0$ systems by the gap between the valence and conduction 
band of (9). This gap is almost constant \cite{rm}
 at equilibrium in the dimerized lattice, with increasing $\delta$ 
offsetting decreasing $\Delta$ in Fig. \ref{epsilon} to give an 
almost constant $\kappa_{el}$. 

 Electronic correlations decrease $\alpha_{el}$ and 
stabilize the I phase of radical ions. CT salts are correlated systems with 
large $U$ in (9). Although their dielectric constants are not expected 
to resemble $U=0$ results, the Peierls instability on the N side 
nevertheless ensures a strong vibrational peak where $\alpha_{el}$ 
is monotonic. 
And increasing dimerization still produces a gap in the electronic 
excitations $E_R>0$ in (1) that reduces $\alpha_{el}$ at the N-I interface. 
Some key $U=0$ features are expected to be retained in CT salts.

\section{Conclusions}

Modeling the  dielectric behavior of molecular materials made up of 
strongly polarizable molecules is challenging in several respects. 
Even in traditional molecular crystals where the overlap 
 of charge densities on different molecules can be safely
neglected, the self-consistent interplay between local electric field
and molecular polarizabilities (and hyperpolarizabilities) is responsible
for the appearance of large collective effects: the response of the 
material is quantitatively, and possibly even qualitatively 
different from the behavior of a collection of non-interacting
molecules. In Section 2 a recent approach to the calculation of 
the dielectric constant of molecular crystals and films
made up of large $\pi$-conjugated  and non-polar molecules
is summarized. The approach is based on the mf approximation, so
that the problem for $N$ interacting molecules reduces to 
the self-consistent problem of a single molecule experiencing the
local electric field generated by the surrounding molecules.
The new approach to electronic polarization also 
yield the polarization energies of cations, anions, and CT states, 
and it is applicable to crystalline thin films.

Even larger effects are expected in materials made up of polar and 
polarizable molecules, such as push-pull chromophores.
A simple toy-model for these materials is based on a two-state
description of each molecular unit. Few-state models, while less accurate
than quantum-chemical descriptions, 
yield  a more transparent description of the basic physics
of the system, and, being amenable to exact solution,
offer valuable information on  the validity of approximations
necessarily introduced when handling more complex Hamiltonians.
Results in Section  3 confirm the reliability of mf approximation 
as introduced in Section 2 for linear (and non-linear) electric
susceptibilities of molecular materials with non-overlapping charge
distributions. Exact results for the toy-model
also confirm the importance of collective  effects in 
aggregates of polar-polarizable chromophores, in which supramolecular 
interactions can tune molecular polarity over a wide range.

Few state models are also useful to address  vibrational contributions to
polarizabilities, and particularly to assess the reliability of standard 
approximation schemes including the BO separation of nuclear and electronic
degrees of freedom and the harmonic approximation. 
A reduced on-site basis, and hence models with few local states,
are also a  useful starting point to face the challenging problem of
polarizability in covalent insulators, such as  CT salts.
 Electronic (charge) instabilities
and lattice instabilities coexist in these materials and compete and/or
 cooperate in complex and interesting ways. 
The peak in the dielectric constant of several CT salts is due to 
the Peierls mode and provides another example of intimate coupling
 between electronic and vibrational degrees of freedom.

\acknowledgments
The work on vibrational contributions to polarizabilities and
 hyperpolarizabilities  in molecular materials and in CT crystals 
was perfomed in collaboration with Luca del Freo. 
The self-consistent treatment of molecular crystals was done 
in collaboration with Eugene Tsiper. We also thank colleagues for many 
discussions: Alberto Girlando 
 in Parma, and Jessica Sin, Robert Pascal, Jr., and Antoine Kahn in Princeton. 
Work in Parma was supported by  Ministero Istruzione Universita' e
Ricerca (COFIN-2001), and by INSTM (PRISMA-2002). 
 Work at Princeton was partially supported by the National 
Science Foundation through the MRSEC program under DMR-9400632.

\vfill
\eject

\begin{table}

\begin{tabular}{|l|c|c|c|c|}
\hline
Molecular Inputs & $\kappa_{11}$ & $\kappa_{bb}$ &
$\kappa_{33}$ & $\theta$ \\ \hline
Expt. (Ref.~\cite{karl}) & 2.49(10) & 3.07(10) & 4.04(20) & 28(2)$^\circ$ \\
TZVP-FIP (Ref.~\cite{reis}) & 2.35 & 3.09 & 4.31 & 31.8$^\circ$ \\
B3LYP/6-311++G(d,p) & 2.23 & 2.91 & 4.03 & 31.6$^\circ$ \\
Expt. gas-phase $\alpha$ & 2.69 & 3.14 & 3.39 & 31.6$^\circ$ \\
Oriented gas & 1.36 & 2.39 & 3.90 &  \\ 
\hline
\end{tabular}
\caption{Principal components of the dielectric tensor and indices of
  refraction, $\kappa =n^2$, of crystalline anthracene. The
  calculated $\kappa$ are based on the indicated molecular
  $\alpha$. The unique axis is $\kappa_{bb}$; $\kappa_{11}$,
  $\kappa_{33}$ are in the $ac$ plane and $\theta$ is the angle
  between $\kappa_{11}$ and $a$.}
\label{tab}
\end{table}

\begin{figure}
\begin{center}
\includegraphics* [scale=1.0] {stack.eps}
\end{center}
\caption{A schematic view of mixed regular and dimerized stacks.}
\label{stack}
\end{figure}

\begin{figure}
\begin{center}
\includegraphics* [scale=1.0] {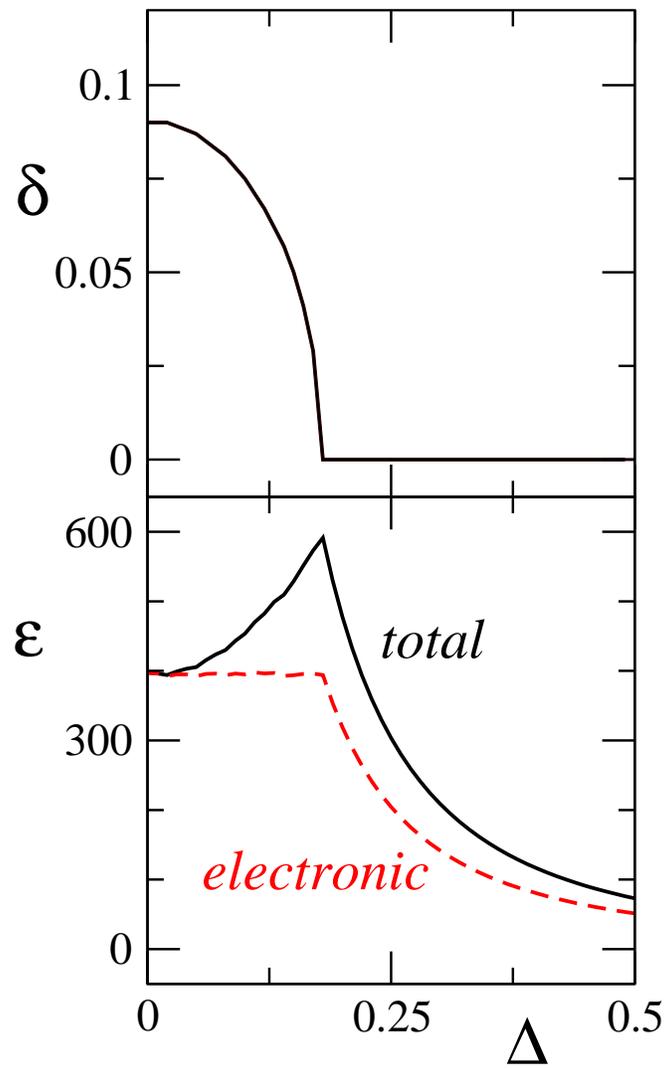}
\end{center}
\caption{The equilibrium dimerization amplitude $\delta$ and
the electronic and total dielectric constants calculated for
the hamiltonian in Eq.~(\ref{hamct}) with $U=0$ and $\epsilon_d=0.28$.}
\label{epsilon}
\end{figure}
  
 \end{document}